\documentclass{article}
\usepackage{ismir,amsmath,cite,url}
\usepackage{graphicx}
\usepackage{color}
\usepackage[nolist,nohyperlinks]{acronym}
\usepackage{amsmath,amssymb,amsthm,mathrsfs,amsfonts,dsfont} 
\usepackage{subcaption}
\usepackage{caption}
\usepackage{soul}
\usepackage[utf8]{inputenc}

\title{Disambiguating Music Artists at Scale with Audio Metric Learning}

\multauthor
{Jimena Royo-Letelier \qquad Romain Hennequin \qquad Viet-Anh Tran \qquad Manuel Moussallam} {Deezer, 12 rue d'Ath\`{e}nes, 75009 Paris, France \\
{\tt\small research@deezer.com}
}
\def\authorname{First author, Second author, Third author, Fourth author}

\begin{document}

\begin{acronym}
    \acro{MSD}{Million Song Dataset}
    \acro{MFCC}{Mel-Frequency Cepstral Coefficients}
    \acro{KNN}{K-Nearest Neighbors}
    \acro{ARI}{Adjusted Rand Index}
    \acro{AMI}{Adjusted Mutual Information}
    \acro{EER}{Equal Error Rate}
\end{acronym}

\maketitle
\begin{abstract}

%Based on a new dedicated homonym artist dataset, we present a new artist clustering task to help disambiguating real-world large scale music catalogs. We subsequently explore the use of metric learning techniques to generate artist embeddings from audio, which is more suitable when sufficient audio data is available during training. \textcolor{black}{We also propose a promising way to improve classical   metric learning techniques that takes advantage of side information, namely artist genres, at training.}      
We address the problem of disambiguating large scale catalogs through the definition of an unknown artist clustering task. We explore the use of metric learning techniques to learn artist embeddings directly from audio, and using a dedicated homonym artists dataset, we compare our method with a recent approach that learn similar embeddings using artist classifiers. While both systems have the ability to disambiguate unknown artists relying exclusively on audio, we show that our system is more suitable in the case when enough audio data is available for each artist in the train dataset. \textcolor{black}{We also propose a new negative sampling method for metric learning that takes advantage of side information such as music genre during the learning phase and shows promising results for the artist clustering task.}	 
\end{abstract}

\section{Introduction}\label{sec:introduction}
\label{sec:intro}
\subsection{Motivation}
\indent With contemporary on-line music catalogs typically proposing dozens of millions of recordings, a major problem is the lack of an universal and reliable mean to identify music artists. Contrarily to albums' and tracks' ISRC\footnote{http://isrc.ifpi.org}, and despite some initiative such as ISNI\footnote{http://www.isni.org}, there exist no unique standardized identifier for artists in the industry. As a direct consequence, the name of an artist remains its de-facto identifier in practice although it results in common ambiguity issues. For example, name artist collisions (e.g.\ \emph{Bill Evans} is the name of a jazz pianist but also the name of a jazz saxophonist and the name of a blackgrass banjo player) or artist aliases (e.g.\ \emph{Youssou N'Dour} vs. \emph{Youssou Ndour}, \emph{Simon \& Garfunkel} vs \emph{Paul Simon and Art Garfunkel}, \emph{Cat Stevens} vs \emph{Yusuf Islam}) are usual. Relying on human resources to clean or verify all artists in the database is practically impossible, although a major issue resulting from these ambiguities is the difficulty to correctly credit artists for their work and the confusion that may arise for end users while exploring catalogs.% (e.g. through search engines). 

Automatically distinguishing between artists is a complicated task, even for human specialists, since there is no one to one relation between a track and an artist. Tracks can be performed by several artists (e.g. \textit{duets} and \textit{featurings}). Albums may contain tracks from different artists (e.g. in \textit{compilations}). Even artist denominations may drastically evolve during their careers. %Moreover, there are a variety of relationships that can exist between a track and an artist (e.g. performer, producer, lyricist, composer etc.).
In this work, %we are not interested in distinguishing between those roles. Our 
our goal is, given a set of recordings, to find a partition of this set for which all tracks from a given subset are associated to the same artist.

%The notion of an artist associated with a recording, while ambiguous, could be roughly defined as the main entity performing a piece of music, which could be a singer, an instrumentalist or an ensemble (e.g. bands or orchestras). In large music catalogs, artist ambiguities such as different artists sharing the same name or one artist with several distinct names are common. Automatically distinguishing artists from tracks or albums is critical when dealing with large music databases of millions of artists, since these are  constantly subject to ambiguous artist naming. For instance, name artist collisions (e.g.\ \emph{Bill Evans} is the name of a jazz pianist but also the name of a jazz saxophonist and the name of a blackgrass banjo player) or unclear artist names (e.g.\ \emph{Youssou N'Dour} vs. \emph{Youssou Ndour}, \emph{Simon \& Garfunkel} vs \emph{Paul Simon and Art Garfunkel}, \emph{Cat Stevens} vs \emph{Yusuf Islam}) are usual. Metadata may not always be sufficient to resolve these ambiguities, especially for unpopular artists with few of them. Moreover, manual curation comes at a high cost since the amount of data is considerable, with on-line music catalogs typically proposing dozens of millions of recordings. %Relying on human resources to clean or verify all artists and albums in the database is practically impossible.

%However, we show in this work that learning embeddings from audio that are useful for artist disambiguation is possible, even if large train databases probably suffer of the previously mentioned problems.  \\ 

\indent In the MIR literature, problems dealing with artist assignation from a recording are most of the time addressed as a classification problem \cite{anchorspace,artistclassification_timbrebased,artistclassification_timbral}, where a set of predefined artists is known. This is not a real case scenario for evolving large music catalogs, since the number of artists can be huge (several millions) and new artists are added every day. In this paper, we propose a new task of unknown artists clustering from audio, without having any ground truth data about the identities of the artist nor a prior information about the number of different artists present in a cluster. To the best of our knowledge, there are no prior work addressing this. Disambiguating homonym artists is a practical application of this task, where a set of tracks must be split into an unknown number of clusters, each corresponding to a different artist entity. We believe that accurately solving this task could results in a major improvement in the quality of large sized catalogs.

To tackle this new task, we propose to use metric learning methods to train a system that outputs artist embeddings from audio. Indeed, as we will explain later, metric learning objective function is primarily designed to ensure that embeddings of samples from the same entity are clustered together. Metric learning also offers the interesting possibility of controlling what an embedding system learns by means of the sampling necessary to feed its loss. 
\textcolor{black}{Here we also suggest to leverage musical relationships among audio tracks as source of information to strengthen the representation learning, allowing to incorporate music side information -such as genre, mood or release date- to the training process.}

\subsection{Overview}

The paper is organized as follows. In Section \ref{sec:previous} we expose how this paper relates to prior work. In Section \ref{sec:system}, we detail the metric learning system used to learn artist embeddings from audio. In Section \ref{sec:evaluation}, we introduce the newly proposed artist disambiguation task and the datasets used for experiments. In Section \ref{sec:results}, we show our results and compare to the previous systems.
%We show our results when using genre side information during the embedding learning.  
Finally, we draw conclusions in Section \ref{sec:conclusion}.

\section{Previous works}
\label{sec:previous}

%We suggest a new learning method that allows to incorporate  music side information, such as genre, mood or release date, in order to strength learning. 

In this paper we propose a method to learn artist embeddings from audio. This approach falls in a general category of methods in identity disambiguation problems  that try to learn a parametric map from content samples (for instance face pictures \cite{facenet} or speaker speech recordings \cite{tristounet}) to a metric space, so that same identity samples are closely located and different identity samples are far away. The same idea has been exploited to address item retrieval in large image datasets (cars, birds, on-line products) \cite{songCVPR16, angular} and to learn audio representations for sound event classification \cite{jansen_towards_2017}. 

For music artist, the embedding approach has been addressed previously in \cite{anchorspace}, where a representation space is constructed using the output probabilities of a multi-class classification system with \ac{MFCC} as input. The space is only used to address the classification of known artists, but would also be suitable to unknown ones. Convolutional deep belief networks were used to learn features that were afterwards used for artist classification \cite{Ng_NIPS2009}: while the evaluation is done on only four artists, the approach learns a representation from unlabeled data which can generalize to unknown artists. In \cite{weston} the authors train a linear system that attempts to capture the semantic similarities between items in a large database by modeling audio, artist names and tags in a single space. The system is trained with multi-task ranking losses, which highly resembles  metric learning methods: each ranking loss takes as input triplets of samples from possibly different kind of sources. Although this approach is very promising, both for the objective function and the use of side information, the same artists are used for train and evaluation. Unfortunately, direct comparison is hard since little details are given about how datasets are obtained. In \cite{park}, artist embeddings are learned using 1d convolutional neural networks trained as mono-label classifier used afterwards as general features extractors. Their approach is able to deal with artists not seen during the training phase. Information is given on how train databases are obtained from the \ac{MSD} \cite{MSD} using the \textit{artist7}\footnote{http://developer.7digital.com} labels.

\indent Several other works address directly the artist classification problem. The current state of the art is inspired by speaker recognition system and makes use of I-vectors to separate artists \cite{artistidentification_noiserobust,artistclassification_timbrebased,artistclassification_timbral}. In \cite{fast_scalable_artist_identification}, an artist fingerprint based on \ac{MFCC} is proposed to tackle the problem of retrieving an artist from an audio track at scale. In \cite{efficient_multivariate_sequence_classification}, multivariate similarity kernel methods are proposed to tackle (among other tasks) artist classification. 
In \cite{artistidentification_sparsemodeling}, the authors focus on the main vocalist, and then vocal separation is used as a  preprocessing for artist classification. In \cite{artistidentification_multimodal}, a multi-modal approach taking advantage of both lyrics and audio is proposed to perform artist classification. In \cite{dieleman2011}, the authors use a convolutional neural network to perform artist recognition on a $50$ artists dataset. While the techniques employed in these works are of interest for their potential use in extracting representations of unknown artists, they usually only consider at the classification of known artists and give no results on the generalization to new artist not seen during training phase, nor address the extraction of representations useful for unknown artists.

%We remark that none of the previous mentioned systems address the clustering of unknown artists, although from the verification task addressed in \cite{park} a clustering can be obtained. 

\section{Proposed method}
\label{sec:system}

\subsection{Metric learning}

The main idea in metric learning is to learn a metric preserving map $f$ from one metric space into another. Using a discrete metric in the first space, this can be exploited to learn embedding spaces where distances correspond with membership to some category. Here we use the discrete metric defined by artist membership in the space of musical audio recordings.

In this membership context, the learning of $f$ can be achieved using the triplet loss mechanism \cite{lmnn}. This relies on using triplets $X = (x_a, x_+, x_-)$, where $(x_a, x_+)$ is a positive pair (samples with same artist membership) and  $(x_a, x_-)$ is a negative pair (samples with different artist membership). The triplet loss $\ell$ contains a term that tries to bring closer the images by $f$ of samples in a positive pairs, and another term that tries to separate the images by $f$ of samples in a negative pair. It writes

\begin{equation}
\label{tl}
   \small \ell(X) = \left|  \|f(x_a) - f(  x_+)\|^2_2 - \|f(x_a) - f(  x_-)\|^2_2  + \alpha \right|_+
\end{equation}

\noindent where $|s|_+ = \max{(0, s)}$. The parameter $\alpha$ here before is a positive constant used to ensure a margin between distances of points from positive and negative pairs. In order to prevent the system to tend to unsuitable states, where all points are mapped to $0$, or where the map $f$ takes arbitrary large values, we constraint the embedding to lie in the unit sphere by imposing $\|f(x)\|_2 = 1$.

\subsection{Training and sampling strategies}
\label{sec:training} 

%\subsubsection{Training}

We train our system using Stochastic Gradient Descent over batches of triplets. Since optimizing over triplets that are already well separated by the system (i.e. $\ell(X) = 0$) is unnecessary and costly, the system is fed only with triplets that are not yet separated. These are called \textit{hard} triplets if the value of 
\begin{equation*}
    \|f(x_a) - f(  x_+)\|^2_2 - \|f(x_a) - f(  x_-)\|^2_2
\end{equation*}
is negative, and \textit{semi-hard} triplets if this value is in $[0, \alpha]$. We set $\alpha = 0.2$ as in \cite{facenet, tristounet}. Notice that during triplets selection process the $i$-th state of the system is used to compute embeddings to filter data for the ($i+1$)-th parameters update.

%\subsubsection{Triplet sampling}

At each iteration, $n$ samples are chosen from $N$ different artists to form positive pairs. For each positive pair one negative sample from the $N-1$ left artists is taken to form a triplet. This leave us with $Nn(n-1)/2$ triplets per iteration that are given to the system for optimization only if they are labeled as hard or semi-hard. 

Notice from the expression (\ref{tl}) that the gradients of $\ell$ are close to zero when the system maps all entries to very close points in the space, so in practice learning can fail with the system being stuck in a ``collapsed'' state where $f(x) = \hat{f}$ for every $x$. We  observe in experiments that correct triplet sampling strategies are crucial to avoid this phenomenon: taking large batches and enforcing the presence of as many different artists as possible in each batch prevents the system from collapsing. 

%\subsubsection{Side information negative sampling}
\textcolor{black}{In order to strengthen the artist representations learned we propose to make use of side information related to music artists. Suppose that we are given tags for artists of the training database. The fact that two artists have a same tag $t$, indicates that these artists share some characteristic. If we want our system to distinguish between similar but not equal artists, an interesting possibility is to train the system to not rely on these characteristics. To implement this idea, we define a probability $p$ to prefer a negative sample $x_-$ with the same tag as the anchor sample $x_a$ when creating a triplet $X$. This is done at each iteration after the first sampling of the $Nn(n-1)/2$ triplets. If for some anchor $x_a$ there is no negative sample $x_-$ with a different tag, or if we do not dispose of tags for the anchor sample, then we fallback to the previous triplet method creation, that is, we choose as negative any sample from another artist. This allows us to make use of side information even if some may be missing in the database, setting thus a flexible sampling framework.}

\section{Experiments and Evaluation}
\label{sec:evaluation}

In this section, we first present our main artist clustering task, the two auxiliary tasks that we use to compare to previous works and validate our metric learning approach, and the datasets used for evaluation. Then, we describe the architecture of the neural network that we use to learn artist representations. Finally, we detail the datasets used during the training of the systems. 

\subsection{Tasks and Evaluations}

\indent The systems studied in this paper output vector embeddings from audio excerpts. In order to perform evaluations, we create track-level embeddings by simply averaging embeddings over $10$ linearly spaced segments of the same track. For the metric learning based system, we also project back the mean embeddings on the unit sphere. 

\subsubsection{Artist classification and verification}

$\,$ \indent \textit{Dataset:} evaluation is made over a dataset of $467$ artists not seen during training of the embedding systems. These artists are taken from the \ac{MSD} as explained in Section \ref{sec:msd}. For each artist we extract $20$ tracks, $15$ tracks are used as ground truth to build artist models and we report the results for $5$ tracks as test cases. 

\textit{Classification task:} we  attribute to each test case the artist identity of its nearest neighbor for the euclidean distance among all the ground truth artist models in the embedding space, and we report the classification accuracy obtained with this procedure.

\textit{Verification task:} this is a binary classification task were given any two track-level embeddings $(e_i, e_j)$ of a test case $i$ and artist model $j$, we decide whether they have the same artist membership. This is achieved by thresholding the euclidean distance between the two embeddings. We may do two types of errors in this task: a false positive error when two embeddings from two different artists are incorrectly classified as sharing the same membership, and a false negative error when two embeddings from the same artist are classified as having different memberships. The higher the decision threshold is, the higher the false negative rate (FNR). Respectively, lowering the threshold results in an increase of the false positive rate (FPR). We report the \ac{EER}, i.e.: the value of FPR and FNR for the threshold at which they are equal.

\subsubsection{Homonym artist clustering}

$\,$ \indent  \textit{Dataset:}  we built an homonym artists database gathering artists which share exactly the same name with manual cleaning. This results in a database of $122$ groups of $2$ to $4$ homonym artists ($102$ groups of $2$ artists, $17$ groups of $3$ artists and  $3$ groups of $4$). Each artist has a total number of $2$ to $14$ albums and $8$ to $168$ tracks.

\textit{Task:} the problem to solve is to discriminate between artists that share the same name. From a set of tracks by different artists (with the same name), the task is to retrieve the actual clusters of tracks having the same artist membership. It is worth noting that this task  may be used in a real life scenario since there is no need of previous knowledge of the identities nor the number of artists present in a group of same named artists. 

We use the \ac{ARI} \cite{rand_objective_1971} and the \ac{AMI} \cite{amivinh} to measure the similarity of the obtained clusters with the ground truth clusters given by the artist memberships. We choose these corrected for chance information theoretic measures since the data size is relatively small compared to the number of clusters present therein \cite{amivinh}.

We present the results on a 5-fold cross-validation of the homonym artists datasets. We used an agglomerative hierarchical clustering algorithm \cite{hac} to group tracks from the same named artists. During the linkage phase we use \textit{ward} method \cite{ward} with euclidean metric for calculating the distance between newly formed clusters, and we apply \textit{distance} criterion to get the flat clusters. We select the optimal parameters based only on the \ac{AMI} performance on the development set, given that we observe a strong correlation with the \ac{ARI} performance. Finally we report the \ac{ARI} and the \ac{AMI}, averaged over the test dataset.

\subsection{Embedding systems}

\indent Our embedding map $f$ consists in a convolutional deep neural network that takes features computed from either $3$s or $9$s long audio samples as input $x$ and outputs points in $\mathcal{S}^1 \subset \mathbb{R}^d$. The audio is down-sampled to $22050$ kH and we use a mel-spectrogram as features, with $128$ mel-filters and $46$ ms long Hann window with $50\%$ of overlapping. We apply dynamic compression at the first layer of the network. Temporal pooling is performed inside the network after the convolution and maxpooling layers, by flattening together the stack and features tensor dimensions and then performing a global average pooling in the time dimension. The output of the network is a $d$-dimensional vector with $d=32$ or $d=256$. Further details of the structure of the network are described in Table \ref{table:network}. We use the RMSProp optimizer \cite{rmsprop} with a learning rate of $10^{-3}$, $\rho=0.9$, $\epsilon=10^{-8}$ and no decay. We compare our metric learning based system to the system in \cite{park} since it is the only previous work that deals with unknown artists.  This system computes artist embeddings of dimension $256$ using the last hidden layer of a 1D convolutional neural network trained as an artist classifier. We refer to \cite{park} for further details. Both systems were implemented using Keras \cite{keras} framework with Tensorflow \cite{tensorflow} as back-end.

\begin{table}
\small
\centering
\begin{tabular}{|c|c|c|c|}
  \hline
  \textbf{layer} & \textbf{size-out}  & \textbf{kernel} & \textbf{activation} \\
  \hline
  dyn. comp.     & 1x388x128  & -        & $\log(1 + 10^4 \cdot)$ \\
  conv           & 16x388x128 & 3x3x16,1 & relu \\ 
  maxpool        & 16x194x64 & 3x3x16,2 & - \\   
  conv           & 32x194x64 & 3x3x32,1 & relu \\
  maxpool        & 32x64x21  & 3x3x32,2 & - \\   
  conv           & 64x64x21  & 3x3x64,1 & relu\\
  maxpool        & 64x21x7   & 3x3x64,2 & - \\
  flatten        & 1x64x147  &    -     & - \\
  average pool   & 1x1x147   &    -     & - \\
 g  dropout        & 1x1x147   &    -     & -  \\
  dense          & 1x1x$d$    &    -     & tanh \\
  $\ell_2$ const.& 1x1x$d$    &    -     & $\cdot \times \| \cdot \|_2^{-1}$ \\
  \hline
\end{tabular}
\caption{Neural network map details for $9s$ long audio segments. The output size is described in stacks x  rows x columns. The kernel is specified as rows x columns x nb. filters, stride. The parameter $d$ is the size of the embedding.} %The number of total parameters is $31520$ for $d=32$ and \textcolor{red}{TODO} for $d=256$.
\label{table:network}
\end{table}

\subsection{Datasets for training embedding systems}
For each of the datasets described in this section, audio was provided by \textit{Deezer} music streaming company. At the web address\footnote{https://github.com/deezer/Disambiguating-Music-Artists-at-Scale-with-Audio-Metric-Learning} are provided the files containing the ids for the tracks as well as the artists names.

We build $3$ datasets with different characteristics in terms of available audio data per artist and statistical distribution of tracks per artist. Our main goal is to test on the homonym artists clustering task the embedding systems obtained in different scenarios. One of the datasets is created to match real life catalog conditions. 

\subsubsection{\textit{\ac{MSD} small dataset.}}
\label{sec:msd}

The first dataset is a sub-sampling of the \ac{MSD}. %after a matching of the datasets tracks onto its own tracks using song metadata (album names, artist names and track titles)
The interest of this dataset is to compare the two studied systems when a small amount of audio data is available for each artist. Following \cite{park} for the dataset construction, we use the \textit{7digitalid} artist labels delivered with the \ac{MSD}, but in addition we take care of cleaning the dataset from potential duplicate artists: we drop \textit{MSD}  artist labels  associated  with more than one \textit{7digitalid} artist labels and viceversa ($6.3\%$ and $5.8\%$ of artist labels). %\textcolor{red}{[TO RE-CHECK, ASK ALBIN FOR HIS EXPLANATION]}).
%We first remove all entries of \textit{MSD} artist labels that are associated with more than one \textit{7digitalid} artist label (which results in dropping $6.3\%$ of the \textit{7digitalid} artist labels \textcolor{red}{[TO RE-CHECK)]}). Then, for each of the remaining \textit{7digitalid} artist labels, we keep only the entries associated with the most common \ac{MSD} artist labels (resulting in a further drop of $5.8\%$ of the \textit{7digitalid} artist labels \textcolor{red}{[TO RE-CHECK]}).
From this cleaned dataset we use the \textit{7digitalid} labels to choose a number of artists between 100 and 2000, and then select $17$ tracks for each artist. These are respectively split into $15$ for training and $2$ to perform early stopping. We extract a $30$s-long excerpts for each track (the position of the excerpts were sampled at  random  between  the  beginning  and  the end  of  each  track) that are further subdivided in $3s$ long segments with no overlapping used to feed the system. This results in $7.5$ minutes of audio per artist in the training sets.

\subsubsection{\textit{Balanced dataset.}} The interest of this second dataset is to compare the two studied systems when access is granted to larger amounts of audio. We choose artists among the $1000$ most popular in the music streaming service that provided the audio. We keep artist that have at least $4$ albums with more than $3$ tracks. From the remaining artists, we first choose a number of them between 25 and 600, and then select $1000$ $\,9s$ long samples for each artist linearly spaced in each track. From these samples we pick $1/3$ for training and $1/6$ to perform early stopping. The split is done at the album level, meaning that two tracks from the same album cannot appear in the same split. This results in $100$ minutes of audio per artist for the training set. We remark that this dataset is balanced in terms of number of samples per artist, a common approach to prevent bias towards common classes in classifier systems. Also, the samples are taken from any region of the track, resulting in dataset with more variability than the \textit{MSD small} dataset.

\subsubsection{\textit{Unbalanced dataset.}} 

\begin{figure}[htb]
    \centering
    \includegraphics[width=1\linewidth]{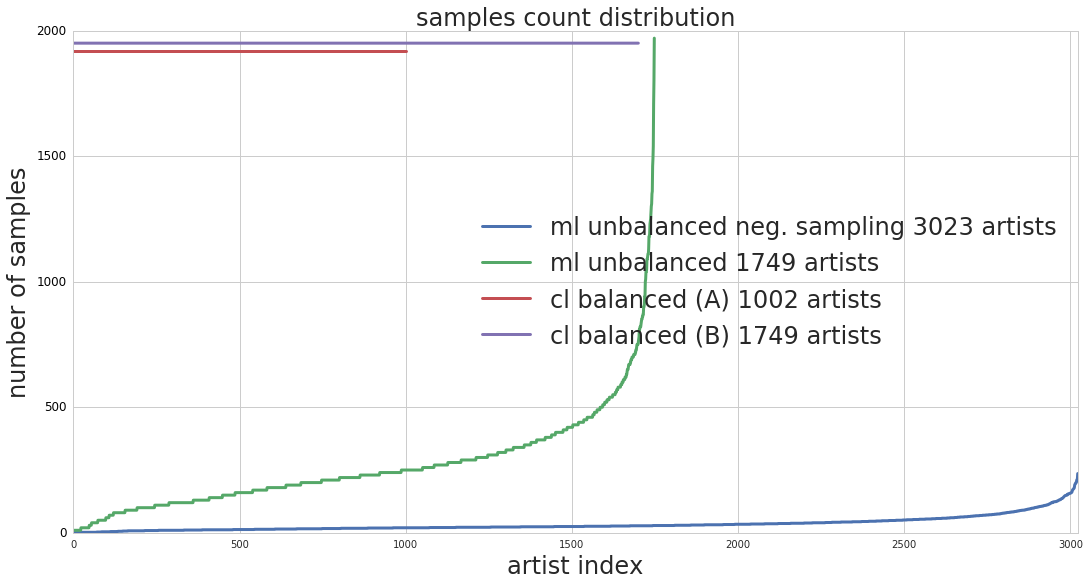}
    \caption{Count distribution of samples by artist in the \textit{unbalanced side information negative sampling} and \textit{unbalanced} datasets for metric learning, and the balanced versions for classifier systems of the \textit{unbalanced} dataset.}
    \label{count4}
\end{figure}

\indent In this third dataset we reproduce the statistical distribution of large music catalogs. We use a match from the Discogs \cite{discogs} dataset onto the music streaming company artist dataset and we keep the genre tags. We drop  artists with less than $4$ tracks. We do not perform any further cleaning, so the resulting dataset is heavily unbalanced and presents typical long tail behavior. Distributions of samples by artist for this dataset are shown in Figure \ref{count4}: the unbalanced dataset exhibit a long-tailed distribution.

%\vspace{-.25cm}
From this dataset we select $10$ $9s$ long samples for each track, which are respectively split into train, evaluation and test. The split is done at the artist level, meaning that two tracks from the same artist cannot appear in the same split. There are $1749$ different artists in the training set, each of one has between $1.5$ and $45$ minutes of audio. 

The interest of this dataset is to study the abilities of each system to make use of all the available audio. Classification systems are usually trained with a balanced dataset. If we have access to a dataset that is not already balanced in terms of classes, we have two options in other to balance it: (A) either cut down samples from the most represented classes or (B) repeat samples of the less represented ones. The former option implies losing data that could have potentially improved the training of the system, while in the second option there is a risk that the classification system over-fit the repeated samples. On the contrary, the triplet sampling of a metric learning system permits to make use of all accessible training data, since the system has the ability of dynamically choosing which data is more relevant for training. To study this, we train the metric learning system with the unbalanced dataset and the classifier system with two balanced versions of it as explained in here before. As we see in Figure \ref{count4}, the (A) dataset contains $1002$ artists and the (B) dataset contains $1749$ artists, each with $2000$ samples. 

\textcolor{black}{Finally,  we study the influence of the new proposed negative sampling method using genre tags with a last \textit{unbalanced side information negative sampling} dataset: from the raw match Discogs we retain 3023 artists, we keep the genre tags and we take only one $9s$ long sample for each track.}

\section{Results}
\label{sec:results}

%We first present in Figure \ref{tsne} a $2$D t-SNE \cite{tsne} visualization of the geometry learnt by the system showing its suitability for clustering of unknown artists.  This demonstrates that the method is able to learn characteristics of artist.

%\begin{figure}[h!]
%    \centering
%    \includegraphics[width=0.75\linewidth]{figs/tnse_art20.png}
%    \caption{\textcolor{red}{TO DO} TSNE}
%    \label{tsne}
%\end{figure}

\begin{figure}[h!]
    \centering
    \includegraphics[width=1\linewidth]{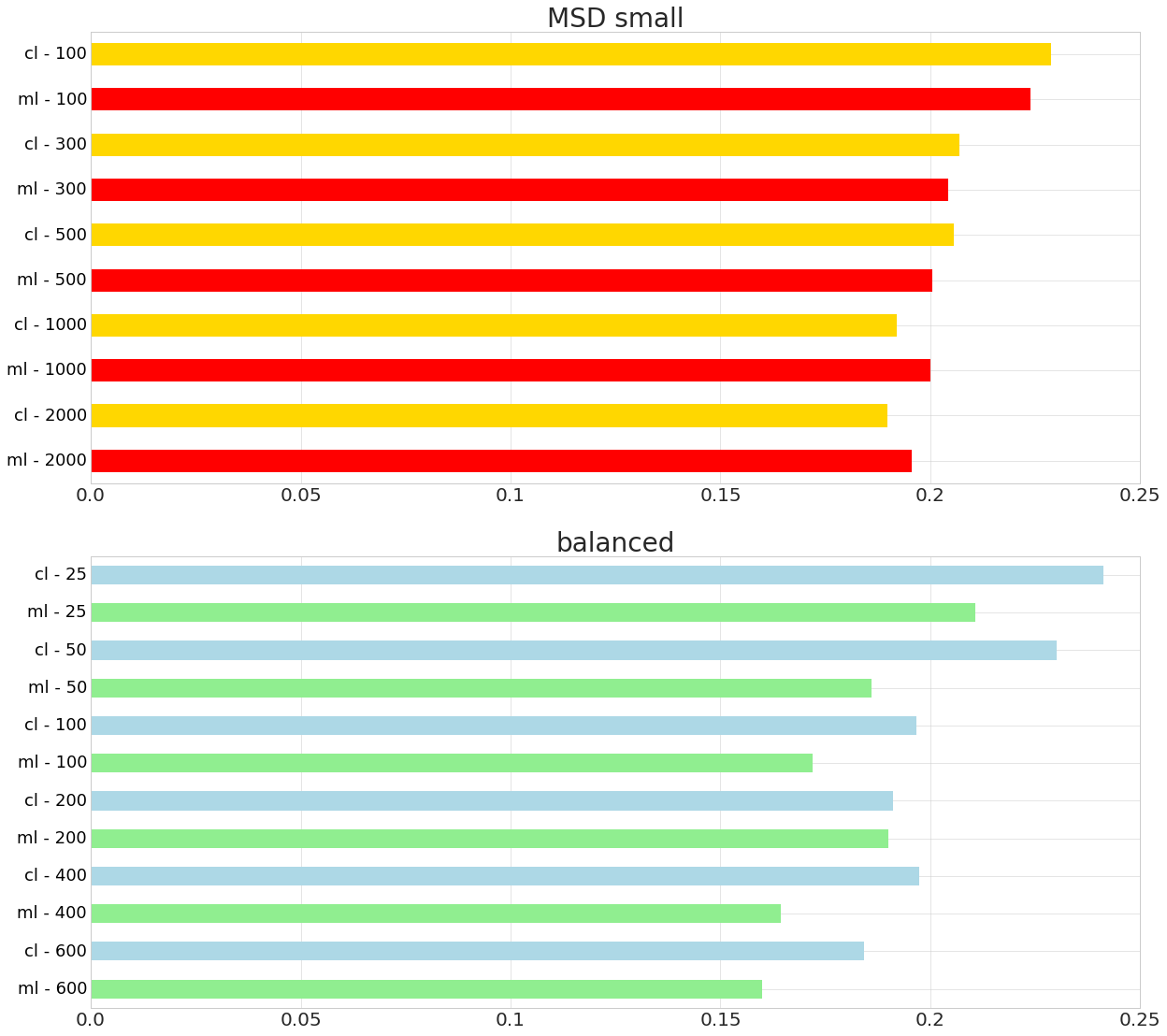}
    \caption{Equal Error Rate results of metric learning (ml - \# artists) and classification (cl - \# artists) embedding systems on the artist verification task for different training datasets and number of artists in the dataset. Lower is better.}
    \label{eer}
\end{figure}

\begin{figure}[h!]
    \centering
    \includegraphics[width=1\linewidth]{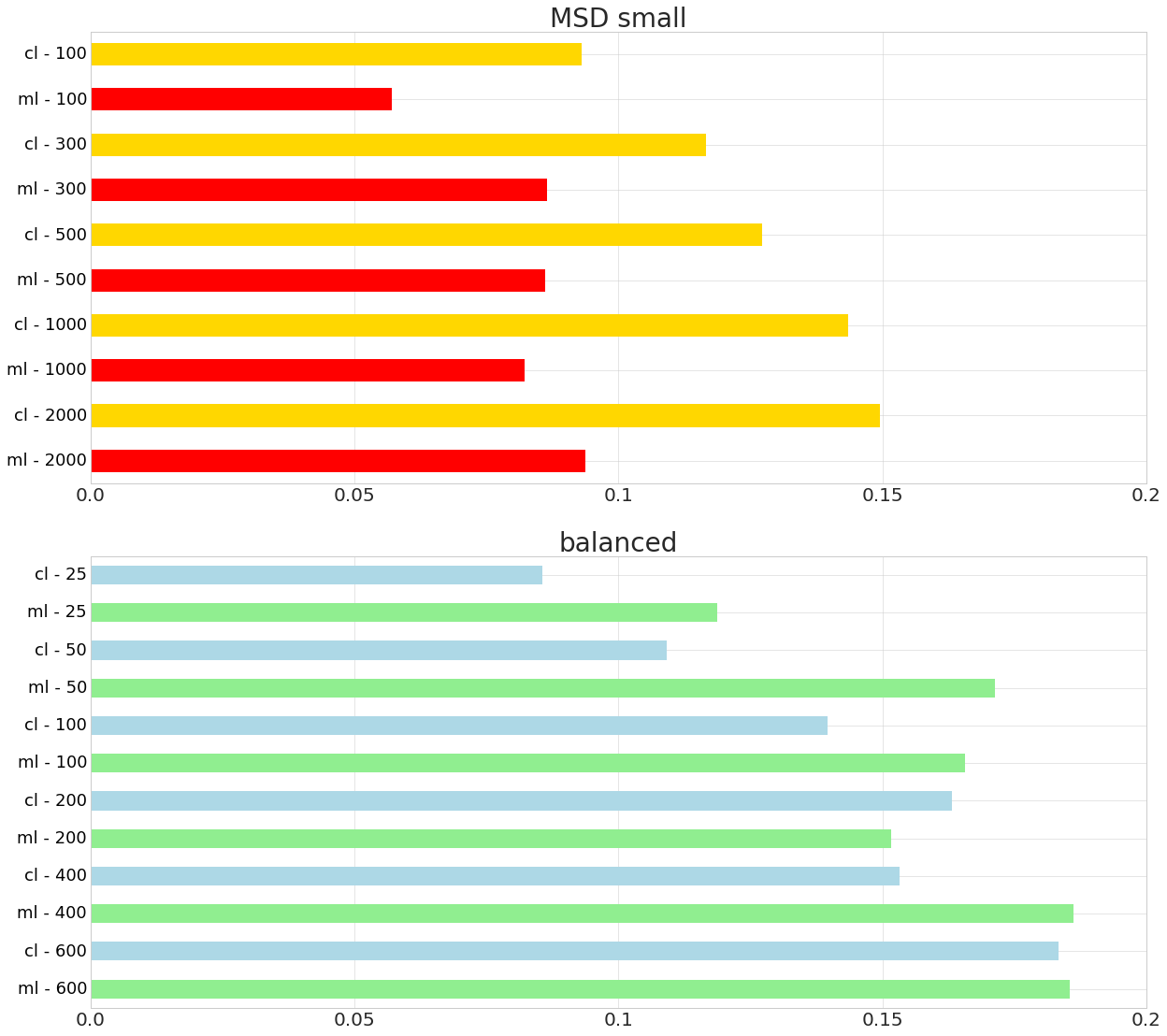}
    \caption{Accuracy results of metric learning (ml - \# artists) and classification (cl - \# artists) embedding systems on the artist verification task for different training datasets and number of artists in the dataset. Higher is better.}
    \label{accuracy}
\end{figure}

We first present in Figures \ref{eer} and \ref{accuracy} the results of the  verification and classification tasks on the \textit{\ac{MSD} small} and \textit{balanced} datasets. We observe that the classifier system performs better than the metric learning based system ($d=256$) when few audio samples are available for each artist (\textit{\ac{MSD} small}), and that inversely, our system outperforms the classifier system when we are provided with larger amounts for each artist (\textit{balanced}). Indeed, metric learning system are generally difficult to optimize, so large quantities of data are needed to make them learn correctly. When this data is accessible for each artist in train datasets, the metric learning system seems to learn better, which validates our approach.

\begin{figure}[h!]
    \centering
    \includegraphics[width=1\linewidth]{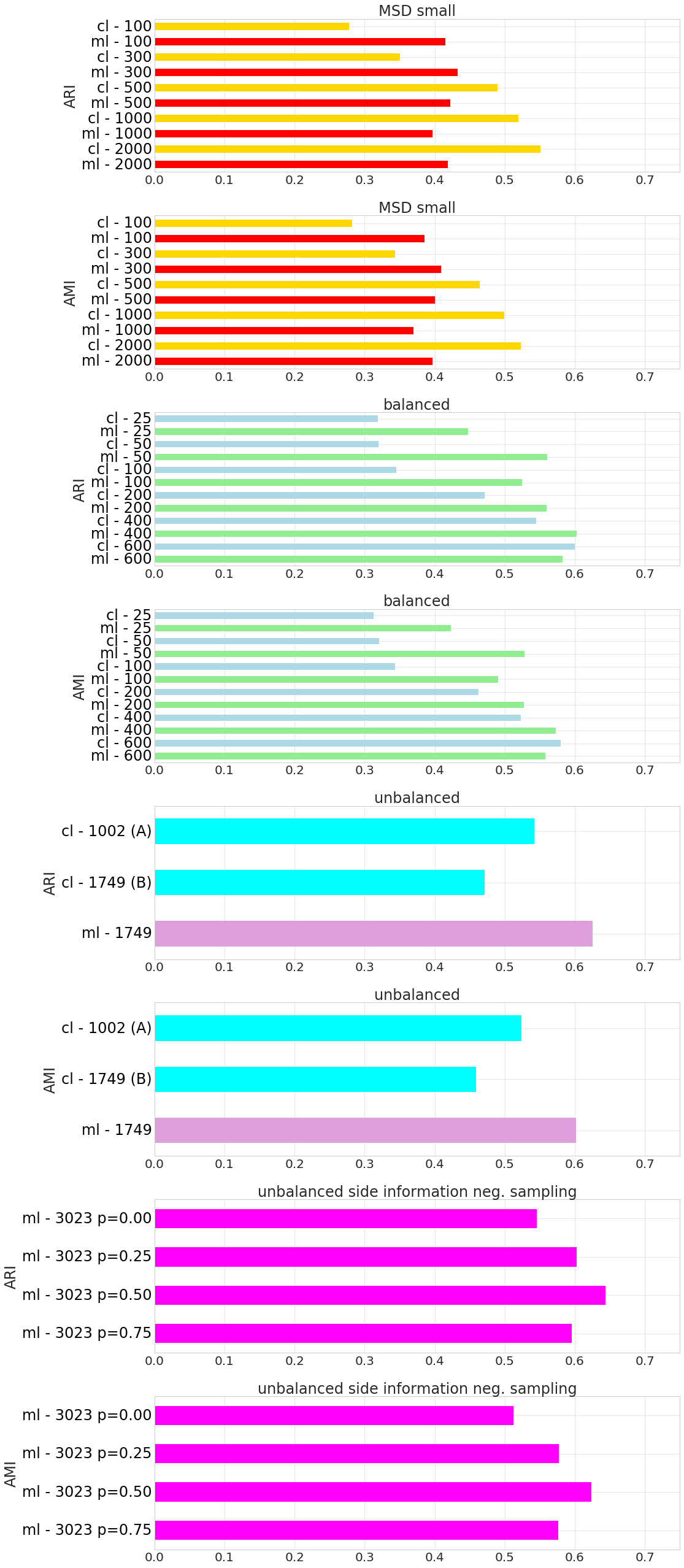}
    \caption{Mean \ac{AMI} and mean \ac{ARI} performances of the metric learning (ml - \# artists) and classification (cl - \# artists) embedding systems on the artist clustering task (5-fold cross-validation) for different training datasets and number of artists in the dataset. Higher is better.}
    \label{hac}
\end{figure}

In Figure \ref{hac} we present the results of the homonym artists clustering task. We first remark that both \ac{ARI} and \ac{AMI} measures take values between $-1$ and $1$, with $0$ as expected value for a random clustering. The obtained results are thus satisfactory, showing the feasibility of the task and making it a compelling candidate to disambiguate unknown artists relying exclusively on audio, for large sized catalogs. 

As we observed for the verification and classification tasks on the \textit{\ac{MSD} small} and \textit{balanced} datasets, the metric learning system generally takes better advantage of larger training datasets. Moreover, the experiments with the \textit{unbalanced} dataset and its balanced versions (\textit{A}) and (\textit{B}) indicate that the metric learning system (d = 32) takes full advantage of all available data, at least when considering the balancing strategies that we proposed. This is of great interest since being able to use all the data in train databases could be beneficial in many other settings.

Finally, we study the results obtained with the side negative sampling strategy that use the genre tags in the \textit{unbalanced side information negative sampling} dataset. We experiment with setting the probability $p$ as explained in Section \ref{sec:training} to different values. Although a linear hierarchy between different values of $p$ is not completely observed, we remark that the best result is obtained with the probability $p=0.5$. These promising results show the potential of using music side information to strengthen the learned artist representations.  
 
  %Moreover, the experiments with the (\textit{unbalanced dataset}) and its balanced versions  (\textit{A}) and (\textit{B}) indicate that the metric learning system ($d=32$) takes better advantage of all available data, at least when considering the balancing strategies that we proposed. This is of great interest since being able to use all the data in train database could be beneficial in many other settings.  
  
\section{CONCLUSION AND FUTURE WORK}
\label{sec:conclusion}

\subsection{Synthesis}

We present a new task of unknown artists clustering to help  disambiguating large scale catalogs, show the interest of it regarding the current problems of artists identification in the music industry, and demonstrate its feasibility with two different artist embeddings methods. Regarding different training datasets conditions (size, amount of audio available, distribution of tracks per artist) one or another could be of better help. We showed that the characteristics of artists learned by the system can generalize to other artists not seen during the learning phase. We prove that metric learning based method is an interesting choice for learning artist representations, in particular by the flexibility of the triplet loss mechanism that allows to better exploit available audio data or to incorporate music side information during training. To this extend, we proposed a new negative sampling method that takes advantage of side information during learning phase and show its relevance when using artist genre tags. 

\subsection{Future work}

An interesting question for subsequent work is to understand the differences between the classification based artist representation and the metric learning based one, and if they learned similar high level characteristics of audio. To this extent a simple concatenation followed by a dimension reduction could be a first solution. More interestingly, we could try to train an embedding system with a linear combination of both losses. Since metric learning loss is difficult to optimize, for instance due to the collapsing problems, classification loss could act as a regularization term. 

Another interesting research direction will be to explore other artist embeddings methods that also can incorporate side information, such as multi-task ranking losses from \cite{weston} or task driven non negative matrix factorization with group constraints as in \cite{serizel}.

Finally, we plan to further investigate the use of side information in negative sampling, and to explore the use of other kinds of sources such as mood or release date. Inversely, an interesting idea would be to investigate how to ameliorate genre representations learning by using artists labels as side information.% in order to cancel the latent factor of belonging to the same artist or to prevent over-fitting artist membership.

\bibliography{ISMIR_artist_disambiguation}

%\section{Acknowledgements}
%The authors would like to thank Robin Vincent for manually checking the Homonym database. The research leading to this work benefited from the WASABI project supported by the French National Research Agency (contract ANR-16-CE23-0017-01).

\end{document}